\documentclass[aps,prl,twocolumn,superscriptaddress]{revtex4-1}

\usepackage{hyperref}
\usepackage{amsmath, mathrsfs, amssymb, wasysym}
\usepackage{graphicx}
\usepackage{color}
\usepackage[caption=false]{subfig}

\begin{document}


\title{Persistent memory in athermal systems in deformable energy landscapes}


\author{Davide Fiocco}
\email{davide.fiocco@epfl.ch}
\affiliation{Institute of Theoretical Physics (ITP), Ecole Polytechnique F\'ed\'erale de Lausanne (EPFL), 1015 Lausanne, Switzerland}
\author{Giuseppe Foffi}
\email{giuseppe.foffi@u-psud.fr}
\affiliation{Institute of Theoretical Physics (ITP), Ecole Polytechnique F\'ed\'erale de Lausanne (EPFL), 1015 Lausanne, Switzerland}
\affiliation{Laboratoire de Physique de Solides, UMR 8502, B\^{a}t. 510, Universit\'e Paris-Sud, F-91405 Orsay, France}
\author{Srikanth Sastry}
\email{sastry@tifrh.res.in}
\affiliation{TIFR Centre for Interdisciplinary Sciences, 21 Brundavan Colony, Narsingi, 500075 Hyderabad, India}
\affiliation{Jawaharlal Nehru Centre for Advanced Scientific Research, Jakkur Campus, 560064 Bangalore, India}


\date{\today}


\begin{abstract}
We show that memory can be encoded in a model amorphous solid subjected to athermal oscillatory shear deformations, and in an analogous spin model with disordered interactions, sharing the feature of a {\it deformable} energy landscape. When these systems are subjected to oscillatory shear deformation, they retain memory of the deformation amplitude imposed in the training phase, when the amplitude is below a ``localization'' threshold. Remarkably, multiple, \emph{persistent}, memories can be stored using such an athermal, noise-free, protocol. The possibility of such memory is shown to be linked to the presence of plastic deformations and associated limit cycles traversed by the system, which exhibit avalanche statistics also seen in related contexts. 
\end{abstract}

\maketitle


Equilibrium in thermodynamic systems is characterized by a loss of memory of previous history, and conversely, systems with broken ergodicity of some form are capable of retaining memory of their past history. Systems displaying order that is induced by a symmetry breaking field can be viewed as a simple example, whereas the memory effects displayed by systems stuck in non-equilibrium, disordered or glassy, states are far more complex \cite{vincent2007ageing,hopfield1982neural,amit1985spin,sethna1993hysteresis,sethna2001crackling,deutsch2003subharmonics,deutsch2004return,reichhardt2012hysteresis}. The presence of a memory of previous history implies that with specific measurements of properties, it is possible to ``read'' such memory. Thus one may speak of training a system to encode specific information, which may be read by making a corresponding measurement later. \\
Recently \cite{keim2011generic,keim2013multiple} it has been showed that a  model~\cite{corte2008random} of a dilute suspension of (non-Brownian) particles contained in a viscous medium, subjected to oscillatory shear-deformation,  can retain memory of the amplitude of  deformation. Repeated oscillations of a given amplitude $\gamma_1$ bring the system toward a stable state in which particles cease to move when viewed stroboscopically, i.e. at zero strain after each cycle. In this sense, we name such a state ``reversible'', following the terminology of \cite{regev2013rheology}.
This procedure encodes a memory that can be read by performing a single cycle of deformation of amplitude $\gamma$ and measuring the fraction of particles $f$ that have moved, as a function of $\gamma$.  The graph of $f$ as a function of $\gamma$ has a kink at $\gamma_{1}$, and for a very large number of training oscillations no particles are displaced by a subsequent reading cycle of amplitude less than $\gamma_{1}$. Moreover, if the training phase consists of alternating oscillations of different amplitudes $\gamma_{1} > \gamma_{2} > \ldots$ the system is capable of showing multiple kinks corresponding to  the training amplitudes $\gamma_{i}$.  For a high number of training cycles, however, the signal of all the $\gamma_{i} < \gamma_{1}$ is suppressed in favor of that of $\gamma_{1}$, so that multiple memories are a \emph{transient} phenomenon. Such multiple transient memories are shown to be stabilized (made ``persistent'', in the language of \cite{keim2011generic}) if noise is introduced in the system in the form of random particle displacements. It was proposed that the phenomenon of such multiple transient memories can be observed in a large variety of systems like granular materials, colloids and  foams, as long as these can (1) reach reversible  states during the initial cyclic training, and (2) there is an ordering of reversible states (so that a state that is reversible under a deformation cycle of amplitude $\gamma_{1}$ is reversible under a cycle of amplitude $\gamma_{2} < \gamma_{1}$). Which systems obey these criteria is a question that remains to be addressed. 
\\
The model studied in \cite{corte2008random, keim2011generic} exhibits a localized/diffusive transition as a function of the strain amplitude, and the memory effects are seen in the localized phase, below a critical threshold $\gamma_{c}$. Remarkably, such a transition, with similar critical qualitative features, is also observed in a model dense amorphous solid~\cite{fiocco2013oscillatory}. In this letter we ask whether memory effects, similar to those seen in Refs.~\onlinecite{keim2011generic, keim2013multiple}, are also present in the case of amorphous solids, something in principle unexpected due to the more complicated nature of their potential energy landscape.
To do so, we study memory effects under oscillatory shear deformation of an amorphous solid which is a binary mixture of particles with Lennard-Jones interactions (BMLJ). We also investigate a disordered spin model (a flavor of the NK model), used in \cite{isner2006generic} to study deformation behaviour of glasses. We find that multiple memories can be encoded in these systems {\it without} noise, and this ability arises from the presence of complex periodic orbits that define the steady state, unlike in \cite{keim2011generic,keim2013multiple}. We also find that discontinuous plastic deformations involved in these cycles exhibit ``avalanche'' statistics also seen in magnetic systems exhibiting memory effects \cite{sethna2001crackling}, earthquakes, and deformations in amorphous solids leading to plastic flow \cite{regev2013rheology,maloney2006amorphous,bailey2007avalanche,lerner2009locality,karmakar2009statistical}. 
\\
The BMLJ samples consist of $N=4000$ Lennard-Jones particles interacting with a Kob-Andersen choice of parameters, cutoff, and composition as in \cite{lacks2004energy, fiocco2013oscillatory}. The density is equal to 1.2 (in reduced units) so that the system is much denser than the suspensions studied in \cite{corte2008random, keim2011generic}.
These are equilibrated at a constant temperature $T=0.466$ via molecular dynamics in the NVT ensemble using LAMMPS \cite{plimpton1995fast}. The equilibrated configurations are then minimized in energy using a conjugate-gradient (CG) algorithm, and the deformation is carried out by means of an athermal-quasi static (AQS) \cite{maloney2006amorphous} procedure where the strain $\gamma_{xy}$ is incremented in steps of $d\gamma_{xy} = 2 \cdot 10^{-4}$ by affinely deforming particle positions, updating boundary conditions and minimizing the energy via CG at each step.
Systems are initially shear deformed by varying the strain between $-\gamma_{1}$ and $\gamma_{1}$ for a certain number of full deformation cycles (the ``training'' phase). Alternatively, we perform the oscillatory training at two amplitudes ($\gamma_{1}$ and $\gamma_{2}$) through a specified repeat sequence. The value of the $\gamma_{i}$ are chosen to be below the critical value $\gamma_{c}$ \cite{fiocco2013oscillatory}, the value under which the system is guaranteed to reach a reversible state for a sufficiently large number of oscillations. After that, samples are subjected to a single cycle of amplitude $\gamma$ (``reading'' phase). 
We monitor the changes in the sample during a reading cycle of amplitude $\gamma$ by measuring the mean squared displacement (MSD) of the particles, averaged over several samples.

\begin{figure}[!ht]
	\includegraphics[width=0.9\columnwidth]{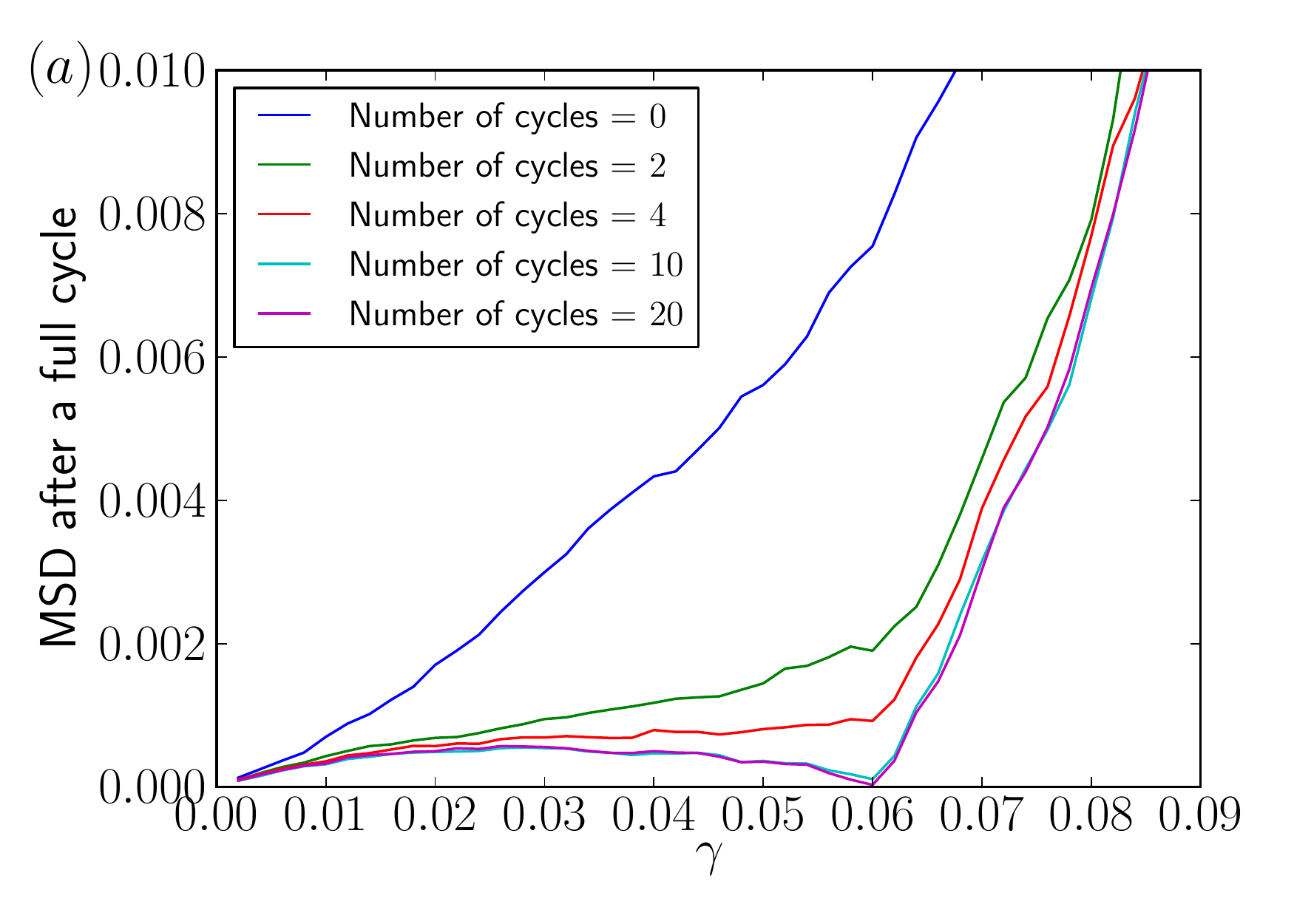}
	\includegraphics[width=0.9\columnwidth]{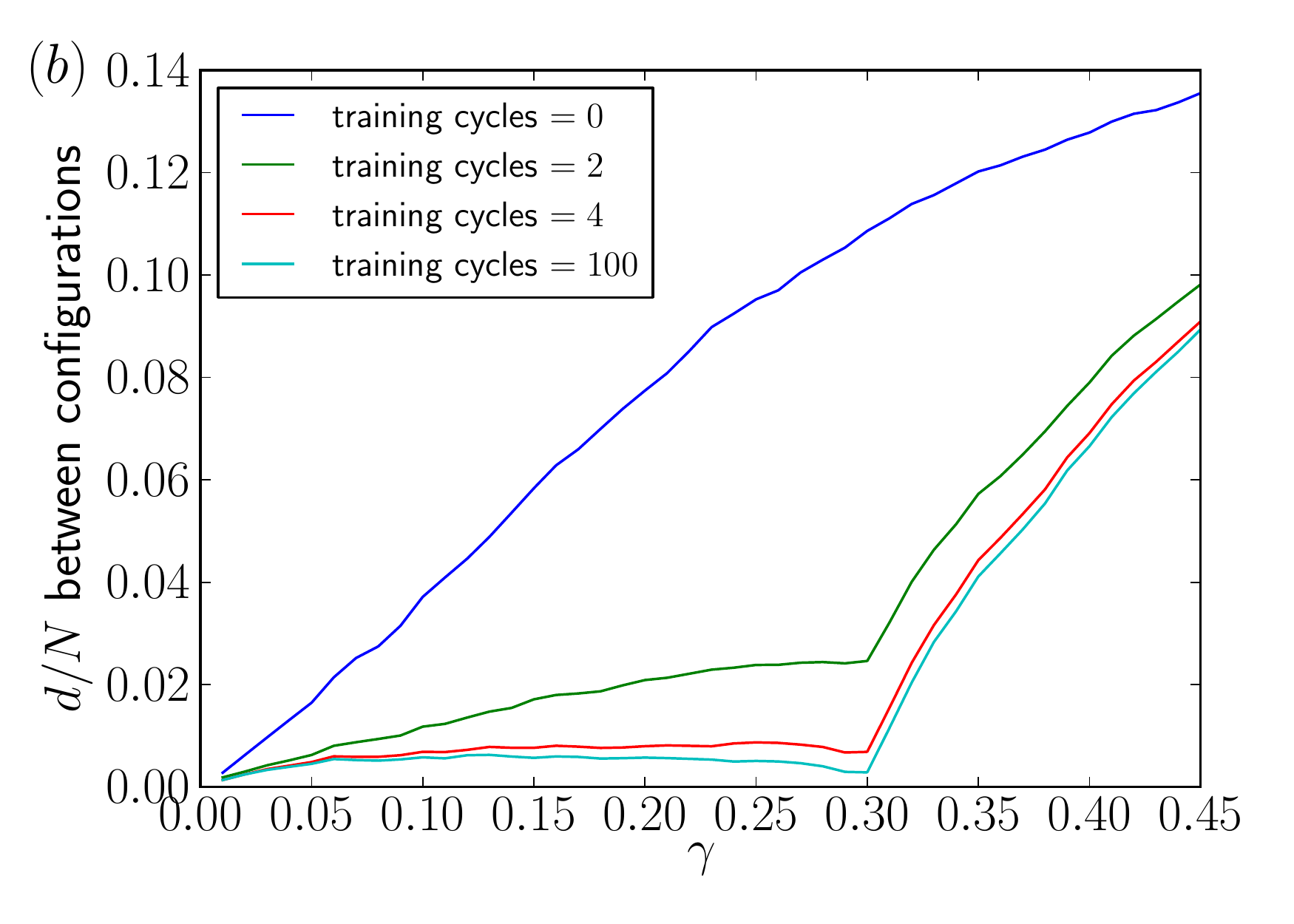}
	\caption{MSD and distance $d$ (scaled by $N$) between configurations before and after a reading cycle as a function of the amplitude $\gamma$, starting from samples trained by oscillatory deformation at (a) $\gamma_1 = 0.06$ for the BMLJ and (b) $\gamma_1 =  0.3$ for the NK model. The value at which the training is performed can be easily read, and configurations can be altered by cycles of amplitude $\gamma < \gamma_{1}$, even if obtained after a long series of training oscillations. \label{fig:MemoryBMLJandNK}}
\end{figure}

\begin{figure}[!ht]
	\includegraphics[width=0.9\columnwidth]{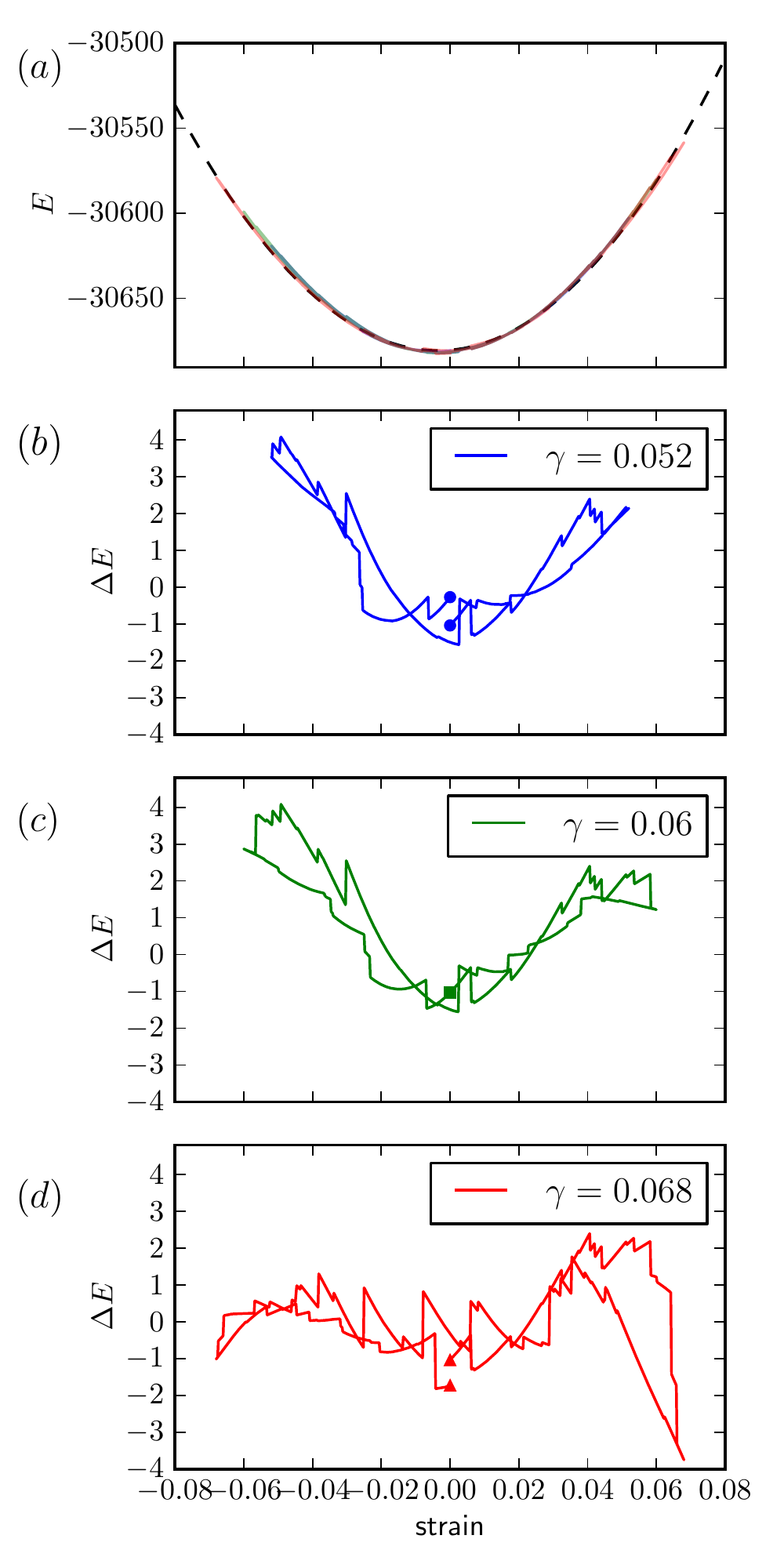}
	\caption{(a) Potential energy measured in a reading cycle starting from a BMLJ sample trained at $\gamma_{1} = 0.06$ for different oscillation amplitudes $\gamma = 0.052, 0.06, 0.068$. The data series almost overlap and are all well fit by the same quadratic profile. In (b-d), the quadratic fitting function is subtracted to obtain $\Delta E$, and the ends of the curves are marked with symbols. The three lines initially follow the same path (for positive strains) and separate as the respective amplitudes are reached. The green line in (c) does join itself at zero strain after a full cycle, but this doesn't happen for the other oscillation amplitudes (the red and blue lines have loose ends) so that samples leave  the stable orbit for $\gamma \neq \gamma_{1}$. \label{fig:ECycleBMLJ}}
\end{figure}

The NK model, on the other side, is a spin model characterized by (an even number) $N$ of lattice sites occupied by spins $m_{i}$ that can take the values $0$ or $1$ with the constraint $\sum_{i} m_{i} = N/2$ (not present in Ref.~\cite{isner2006generic}). 
Each spin has $K$ neighbors $m_{i}^{1}, \ldots, m_{i}^{K}$, and the energy of the system, $E = E(m_{1}, \ldots, m_{N})$, is defined as
\begin{equation}
	E = - \frac{1}{2N} \sum_{i = 1}^{N} \left[ 1 + \sin(2 \pi (a_{i} + \gamma_{NK} b_{i})) \right]
	\label{NKEnergy}
\end{equation}
where $\gamma_{NK}$ is the value of ``shear strain'', and the values $a_{i}$ and $b_{i}$ depend on  the $i$-th spin and its neighbors, i.e. $\{m_{i}, m_{i}^{1}, \ldots, m_{i}^{K}\}$ , according to the maps $a$ and $b$
\begin{align}
	\{0, 1\}^{K+1} & \xrightarrow{a} [-1,1], \\
	\{0, 1\}^{K+1} & \xrightarrow{b} [0,1],
\end{align}
that associate every possible binary $K+1$-tuple to a random value chosen with uniform probability in the intervals written above.
The energy in Eq.~\ref{NKEnergy} implies an energy landscape where the roughness grows with the parameter $K$. The strain parameter $\gamma_{NK}$ changes the energy continuously, and allows one to perform trainings and reads as in the BMLJ case. Two NK configurations are considered neighboring if they are converted to the other by the application of a single Kawasaki exchange move \cite{kawasaki1972phase}. Equilibrated configurations can be obtained by performing a Monte Carlo run at
temperature $T$ using Kawasaki moves. For each of these, the associated inherent structures (local energy minima) are found by
steepest descent with Kawasaki moves and their average energies depend on the equilibration temperature $T$ in qualitative agreement to model glassy systems. We perform oscillatory athermal deformations on NK samples (with $N=20, K = 10$), starting from inherent structures obtained from configurations equilibrated at $T=1$. $\gamma_{NK}$ is incremented in steps $d\gamma_{NK} = 0.005$ and the energy is minimized at each step. Different NK configurations at $\gamma_{NK} = 0$ are compared by measuring their Euclidean distance $d$ divided by $N$. $d^2/N$ is the direct analogue to the MSD of the BMLJ case. \\
Results for different training for the BMLJ and the NK model are presented in Fig.~\ref{fig:MemoryBMLJandNK}. It can be noticed (see Fig.~\ref{fig:MemoryBMLJandNK}a) that the BMLJ samples trained with a maximum amplitude $\gamma_{1}=0.06$ are \emph{not} necessarily stable under cycles of amplitude $\gamma < \gamma_{1}$ since the MSD is not zero for such $\gamma$ and thus \emph{there is no ordering of reversible states}. Fig.~\ref{fig:MemoryBMLJandNK}b shows that the same observation holds also for the NK model. This can be rationalized by the fact that reversibility of given configurations under a full cycle has a completely different mechanism in the BMLJ and NK models compared to the model in Ref.~\cite{keim2011generic}. In the models discussed here, the dynamics is dependent on the evolution of the energy landscape under increasing strain and the system undergoes various inherent structure transitions as a consequence of the destabilization of the energy minima (this, in the BMLJ case, is related to collisions with potential energy saddle points \cite{malandro1999relationships}) during a cycle of amplitude $\gamma_{1}$, whereas this is not at all the case in Ref.~\cite{keim2011generic}.  As indicated in Fig.~\ref{fig:ECycleBMLJ} reversible states are achieved in our case with the rearrangements associated with inherent structure transitions canceling out {\it over the full strain cycle}, and not necessarily with step-by-step reversibility - i.e. the sequence of inherent structure transitions does not need to retrace when strain is reversed in order to produce reversible states.

\begin{figure}[!ht]
	\includegraphics[width=0.9\columnwidth]{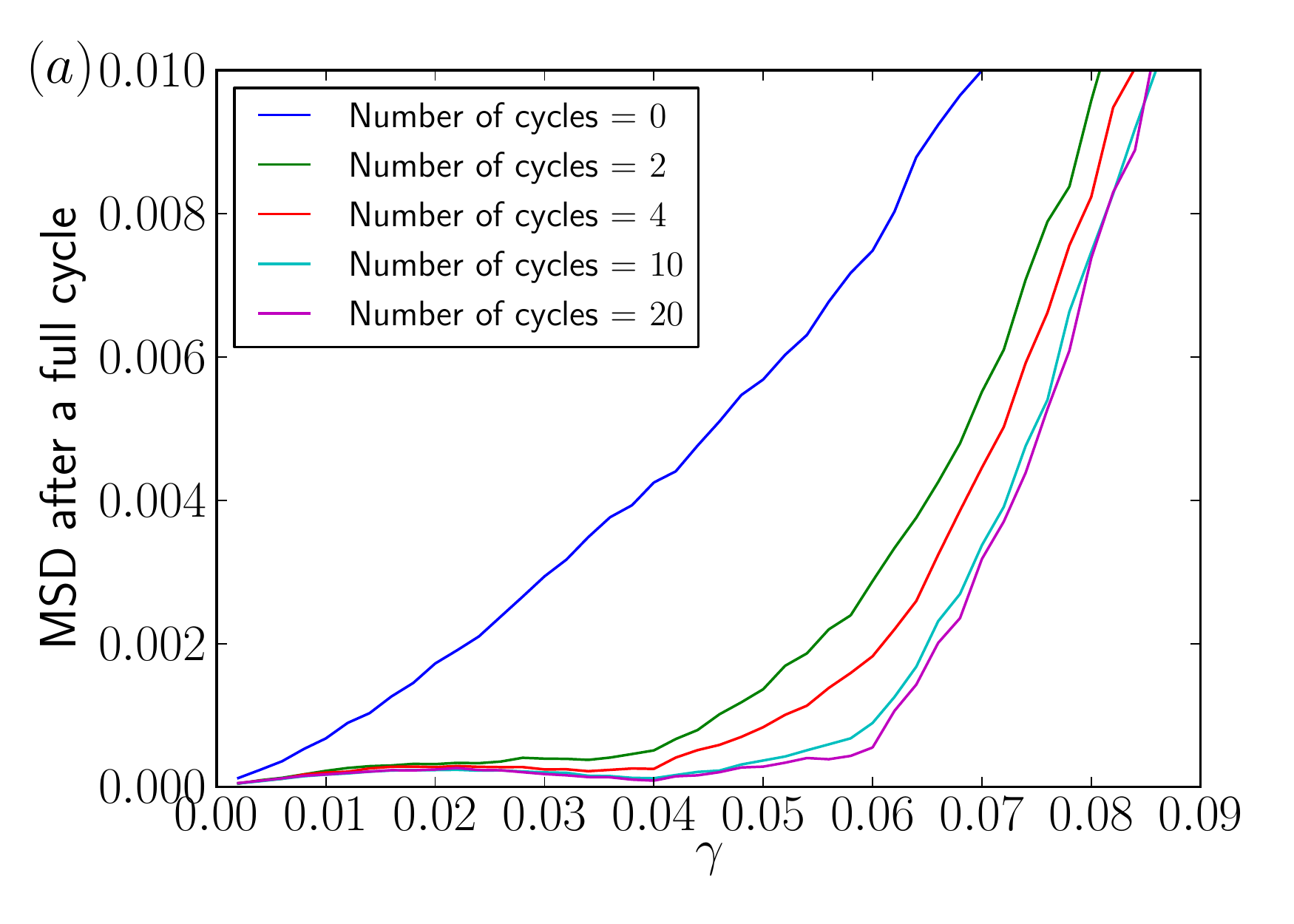}
	\includegraphics[width=0.9\columnwidth]{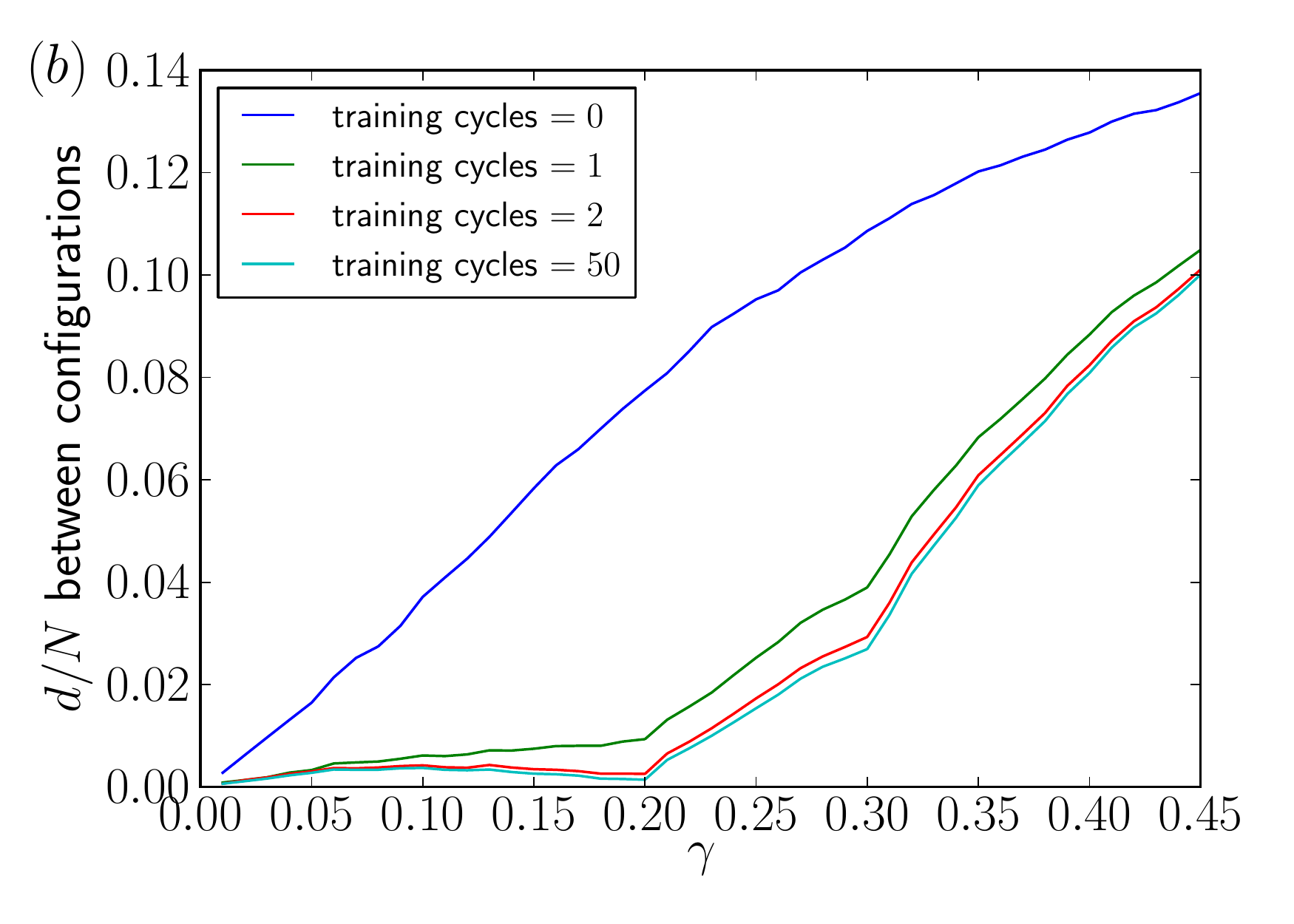}
	\caption{MSD and distance $d$ (scaled by $N$) between configurations before and after a reading cycle as a function of the amplitude $\gamma$, starting from samples trained by oscillatory deformation at (a) $\gamma_1 = 0.06$, $\gamma_2 = 0.04$ (as described in the text) for the BMLJ  and (b) $\gamma_1 =  0.3$, $\gamma_2 = 0.2$ for the NK model. The values at which the training is performed can be easily read even after a large number of oscillations, as samples retain multiple memories of the training phase in a persistent way. \label{fig:DoubleMemoryBMLJandNK}}
\end{figure}

Fig.~\ref{fig:ECycleBMLJ}b-d show the potential energy $\Delta E$ (once the parabolic energy background shown in Fig.~\ref{fig:ECycleBMLJ}a has been subtracted), during reading cycles for different amplitudes, plotted as a function of strain. $\Delta E$ exhibits discontinuities corresponding to inherent structure transitions.
If the amplitude of the reading cycle is equal to the training value $\gamma_1$, $\Delta E$ traces a closed loop (Fig.~\ref{fig:ECycleBMLJ}c). 
If a system tracing such an orbit in configuration space with strain amplitude $\gamma_{1}$ is deformed by a smaller (Fig.~\ref{fig:ECycleBMLJ}b) or higher (Fig.~\ref{fig:ECycleBMLJ}d) amplitude, the sequence of transitions is not necessarily the same as for $\gamma_{1}$ and the system does not return to the initial state. The destabilization at reading amplitudes smaller than $\gamma_1$  is not present in the systems studied in \cite{keim2011generic, corte2008random}, where configurations stable for oscillations of amplitude $\gamma_{1}$ are stable for all $\gamma < \gamma_{1}$ if noise is absent.\\
As seen in Fig.~\ref{fig:ECycleBMLJ}, stable states associated to some value $\gamma_{i}$ can be destabilized (and thus their memory erased) by oscillations of any amplitude $\gamma_{j} \neq \gamma_{i}$. Thus, if a sample is trained by alternating cycles of different amplitudes $\gamma_{i}$ the largest amplitude $\gamma_{1}$ doesn't necessarily take over, \emph{even in the absence of noise}.
This effect is clearly seen in Fig.~\ref{fig:DoubleMemoryBMLJandNK}a for BMLJ systems  subjected to strain cycles of the type $0 \rightarrow \gamma_{1} \rightarrow -\gamma_{1} \rightarrow 0 \rightarrow \gamma_{2} \rightarrow -\gamma_{2} \rightarrow 0$ (with $\gamma_{2} = 0.04,\ \gamma_{1} = 0.06$) in the training phase. 
In this case, for a high number of training cycles, the MSD plotted as a function of $\gamma$ converges to a curve showing kinks at both $\gamma_{1}$ and $\gamma_{2}$.  The information about the two (or in general multiple) training amplitudes is thus encoded and retained for arbitrarily large numbers of training cycles in a persistent manner, as opposed to transiently, as in the absence of noise in Refs.~\onlinecite{keim2011generic, keim2013multiple}.
The deformation at the largest deformation amplitude $\gamma_{1}$ does \emph{not} eventually erase the signal of $\gamma_{2}$ because each of the training oscillations at some amplitude is able to erase part of the information encapsulated by the training at the other amplitudes. \\
Multiple memories are also shown by the NK model when it is deformed with the same protocol followed for the BMLJ. As shown in 
 Fig.~\ref{fig:DoubleMemoryBMLJandNK}b such memories are also persistent.

\begin{figure}[!ht]
	\includegraphics[width=0.7\columnwidth]{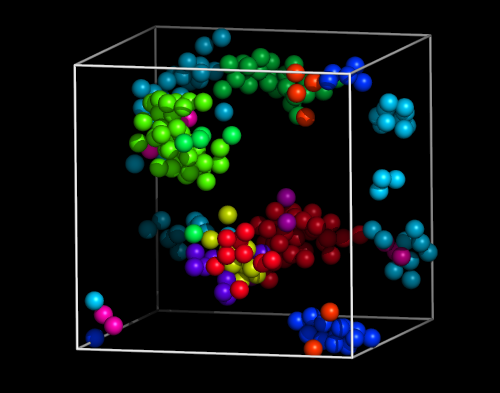}
	\includegraphics[width=0.9\columnwidth]{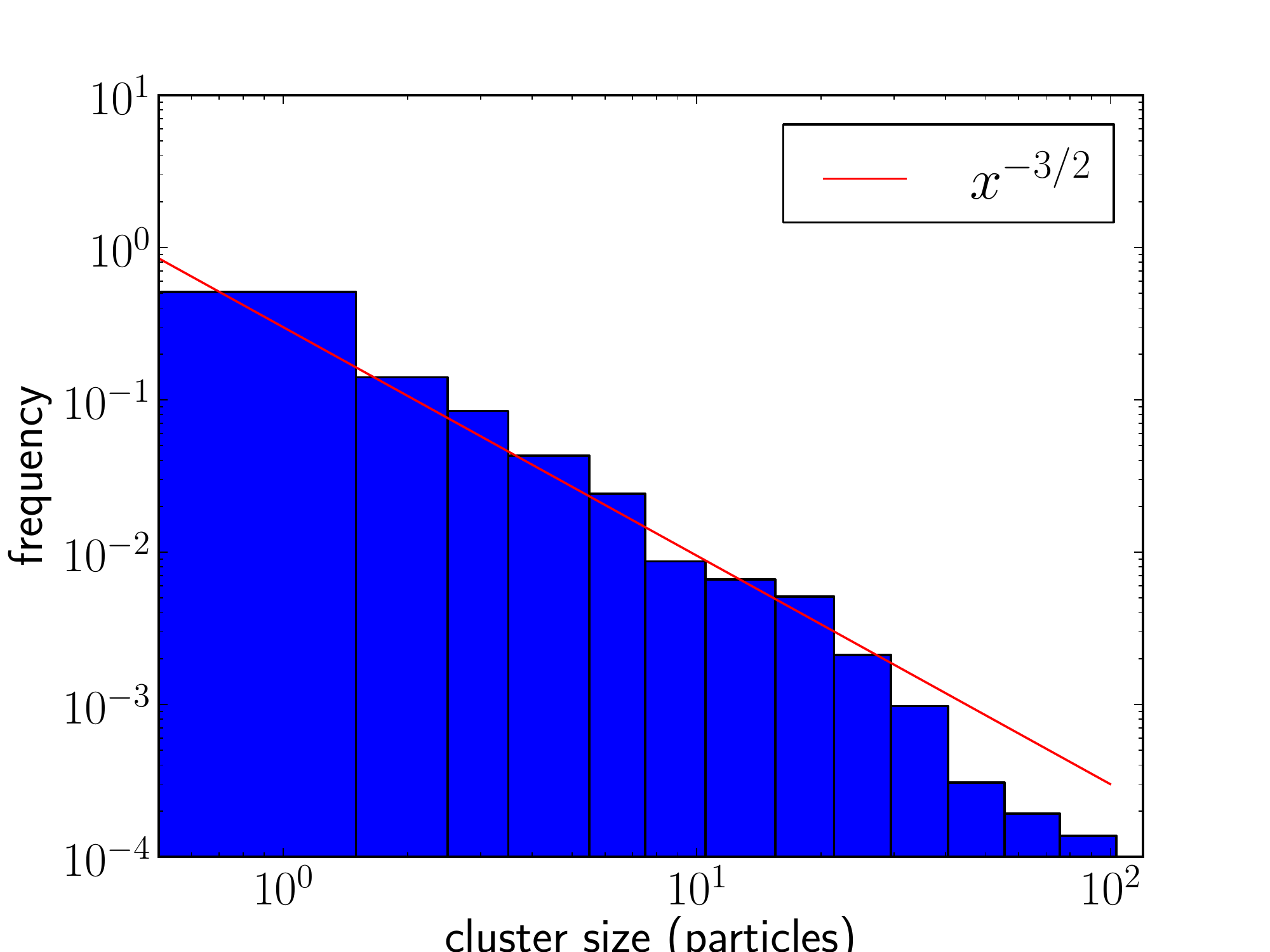}
	\caption{(a) Snapshot of a BMLJ reversible sample trained and read with $\gamma_{1}$ and $\gamma = 0.06$. Particles that move more than $0.1\, \sigma_{AA}$ during different transitions occurring in the reading cycle are drawn in different colors at the positions that they occupy at the beginning of the cycle.
	(b) Plot of the size of clusters of particles that move more than $0.1 \sigma_{AA}$ (particles belong to the same cluster if their distance is $< 1.4\sigma_{AA}$) during transitions. Large clusters become increasingly rare as their size grows.\label{fig:Clusters}}
\end{figure}

To characterize the spatial features of the particle rearrangements that occur during a reading cycle in the trained BMLJ samples, we show in Fig.~\ref{fig:Clusters}a  particles that move more than $0.1\, \sigma_{AA}$ (being $\sigma_{AA}$ the diameter of the largest of the KA components), rendering particles displaced in the same transition with the same color. The choice of the cutoff is non-trivial (see e.g.\ \cite{lerner2009locality} for a discussion); here, we follow the observation in \cite{schroder2000crossover} that particle displacements exhibit a power law distribution arising from elasticity, followed by an exponential tail, and choose the cutoff that separates the two regimes. Particles that move the most in such rearrangements typically do cluster together in space. Typical clusters range from 1 to about 100 particles for our system size, and interestingly, the sizes are distributed according to a power law with exponent $\sim -3/2$ (see Fig.~\ref{fig:Clusters}b), similarly to systems exhibiting avalanches \cite{sethna2001crackling}, and are thus not {\it localized} in any simple way. A better characterization of the statistics of these events demands further analysis of system size and noise/temperature effects, which is beyond the scope of the present work.

To summarize, we have studied memory effects in two model systems (BMLJ and NK) subjected to athermal quasi static deformations, a procedure that is expected to describe the qualitative behaviour of disordered solids at low temperature and low shear rate. These systems evolve to a steady state upon repeated cyclic deformations at a fixed amplitude (below the critical value $\gamma_{c}$ \cite{fiocco2013thesis}), and this ``training'' amplitude can be read by performing a single cycle of strain at varying amplitudes, similarly to the observation for a model of suspensions in Refs.~\onlinecite{keim2011generic, keim2013multiple}. Differently from ~\cite{keim2011generic, keim2013multiple}, however, the systems that we study show no ordering of reversible states, and we have used this property to demonstrate that in these systems it is possible to encode multiple memories that are persistent. This possibility is related to the fact that reversible states attained at the training strain amplitude exhibit non-trivial periodic orbits, which are disrupted by cyclic shear strain at any other amplitude. 
Reading the information encoded in our systems is a destructive operation, and devising protocols whereby memory is tolerant to multiple read cycles poses an interesting challenge.
As verified in \cite{fiocco2013oscillatory}, finite size effects do not affect the qualitative features of the dynamical transition in this system, and we therefore expect size effects not to significantly bear upon our analysis. The displacement events corresponding to transition steps in these periodic orbits are found to be spatially correlated displacements of particles, which however exhibit a broad, power law distribution of sizes. Our observations should be of relevance to memory effects in a wide range of glassy systems subjected to oscillatory external fields. In particular, it will be interesting to explore analogies with disordered spin systems in oscillatory magnetic fields.

We thank A. Dhar, S. Franz, S. Karmakar, J. Kurchan, G. McKenna, S. R. Nagel, E. Vincent and T. Witten for illuminating discussions and F. Varrato for careful reading of the manuscript. We acknowledge support from the Indo-Swiss Joint Research Programme (ISJRP). D.F. and G.F. acknowledge financial support from Swiss National Science Foundation (SNSF) Grants PP0022\_119006 and PP00P2\_140822/1.

\end{document}